\RequirePackage{fix-cm}
\documentclass[smallextended]{svjour3}       
\smartqed  
\usepackage{graphicx}
\usepackage{natbib}
\usepackage{amsmath}
\usepackage{amsfonts}
\usepackage[labelsep=period]{caption}
\usepackage{array}
\usepackage{caption}
\newtheorem{assumption}{Assumption}
\newtheorem{algorithm}{Algorithm}
\usepackage{amssymb}

\begin{document}

\title{Fast algorithms for the quantile regression process
}


\author{Victor Chernozhukov \and Iv\'an Fern\'andez-Val \and Blaise Melly
}


\institute{V. Chernozhukov \at
              Department of Economics, Massachusetts Institute of Technology, Cambridge, USA \\
              \email{vchern@mit.edu}           
           \and
           I. Fern\'andez-Val \at
              Department of Economics, Boston University, Boston MA, USA
              \\
              \email{ivanf@bu.edu}
\and
           B. Melly \at
              Department of Economics, University of Bern, Bern, Switzerland
              \\
              \email{blaise.melly@vwi.unibe.ch}
}

\date{Received: date / Accepted: date}

\maketitle

\begin{abstract}
The widespread use of quantile regression methods depends crucially on the existence of fast algorithms. Despite numerous algorithmic improvements,  the computation time is still non-negligible because researchers often estimate many quantile regressions and use the bootstrap for inference. We suggest two new fast algorithms for the estimation of a sequence of quantile regressions at many quantile indexes. The first algorithm applies the preprocessing idea of \citet{portnoy1997gaussian} but exploits a previously estimated quantile regression to guess the sign of the residuals. This step allows for a reduction of the effective sample size. The second algorithm   starts from a previously estimated quantile regression at a similar quantile index and updates it using a single Newton-Raphson iteration. The first algorithm is exact, while the second is only asymptotically equivalent to the traditional quantile regression estimator. We also apply the preprocessing idea to the bootstrap by using the sample estimates to guess the sign of the residuals in the bootstrap sample. Simulations show that our new algorithms provide very large improvements in computation time without significant (if any) cost in the quality of the estimates. For instance, we divide by 100 the time required to estimate 99 quantile regressions with 20 regressors and 50,000 observations.
\end{abstract}

\section{Introduction}
\label{intro}

The idea of estimating regression parameters by minimizing the sum of absolute errors (i.e. median regression) predates  the least squares method by about 50 years. Roger Joseph Boscovich suggested in 1857 to fit a straight line to observational data by minimizing  the sum of absolute residuals under the restriction that the sum of residuals be zero. Laplace found in 1899 that the solution to this problem is a weighted median.\footnote{Among others, see Chapter 1 in \cite{stigler1986history}.}  Gauss suggested later to use the least squares criterion. For him, the choice of the loss function is an arbitrary convention and ``it cannot be denied that Laplace's convention violates continuity and hence resists analytic treatment, while the results that my convention leads to are distinguished by their wonderful simplicity and generality.'' Thus, it was mostly due to the simplicity of the analysis and of the computation that least squares methods have dominated the statistical and econometric literature during two hundred years.\footnote{See \cite{koenker2000galton} and \cite{koenker2017computational} for a more detailed historical account of the computation of median regression.}

Only the development of the simplex algorithm during the 20th century and its adaptation for median regression by \cite{barrodale1974solution} made least absolute error methods applicable to multivariate regressions  for problems of moderate size. \cite{portnoy1997gaussian} developed an interior points algorithm and a preprocessing step that significantly reduces the computation time for large samples with many covariates. These algorithms are implemented in different programming languages and certainly explain the surge of applications of quantile regression methods during the last 20 years.     

Despite these improvements, the computation time is still non-negligible in two cases: when we are interested in the whole quantile regression process and when we want to use bootstrap for inference. Researchers are rarely interested in only one quantile regression but often estimate many quantile regressions. The most common motivation for using quantile regression is, indeed, to analyze heterogeneity. Some estimation and inferential procedures even require the preliminary estimation of the whole quantile regression process. For instance, \cite{koenker1987estimation} integrate the trimmed quantile regression process to get conditional trimmed means. \cite{Koenker2002}, \cite{ChFe05} and \cite{ChHa06} test functional null hypotheses such as the absence of any effect of a covariate or the location-scale model, which requires estimating the whole quantile regression process. \cite{MachadoMata05} and \cite{chernozhukov2013inference} must compute the whole quantile regression process to estimate the conditional distribution function by inverting the estimated conditional quantile function. In a second step, they integrate the conditional distribution function over the distribution of the covariates to obtain unconditional distributions.

In addition, researchers often use the bootstrap to estimate the standard errors of quantile regression. Many simulation studies show that it often provides the best estimate of the variance of the point estimates. It also has the advantage of avoiding the (explicit) choice of a smoothing parameter, which is required for the analytical estimators of the variance. Finally, it is difficult to avoid the bootstrap (or other simulation methods) to test some types of functional hypotheses.

In these two cases, when the number of observations is large, the computational burden of quantile regression may still be a reason not to use this method. In their survey of the decomposition literature, \cite{fortin2011decomposition} mention the following main limitation of the quantile regression approach: ``This decomposition method is computationally demanding, and becomes quite cumbersome for data sets numbering more than a few thousand observations. Bootstrapping quantile regressions for sizeable number of quantiles $\tau$ ($100$ would be a minimum) is computationally tedious with large data sets.'' Further improvements are clearly needed to enable more widespread applications of quantile regression estimators.

In this paper we suggest two new algorithms to estimate the quantile regression process. The first  algorithm is a natural extension of the preprocessing idea of \cite{portnoy1997gaussian}. The intuition is fairly simple. The fitted quantile regression line interpolates $k$ data points, where $k$ is the number of regressors. Only the sign of the residuals of the other observations matters for the determination of the quantile regression coefficients. If one can guess the sign of some residuals, then these observations do not need to enter into the optimization problem, thus reducing the effective sample size.  When many quantile regressions are estimated, then we can start with a conventional algorithm for the first quantile regression and then progressively climb over the conditional quantile function, using recursively the previous quantile regression as a guess for the next one. This algorithm provides numerically the same estimate as the traditional algorithms because we can check if the guesses were correct and stop only if this is the case.

This algorithm seriously reduces the computation time when we estimate many quantile regressions. When these improvements are still insufficient, we suggest a new estimator for the quantile regression process. The idea consists in approximating the difference between two quantile regression coefficient vectors at two close quantile indexes using a first-order Taylor approximation. The one-step algorithm starts from one quantile regression estimated using one of the existing algorithms and then obtains sequentially all remaining quantile regressions using one-step approximations. This algorithm is extremely quick, but the estimates are numerically different from the estimates obtained using the other algorithms. Nevertheless, the one-step estimator is asymptotically equivalent to the standard quantile regression estimator if the distance between the quantile indexes decreases sufficiently fast with the sample size.

We also apply the preprocessing idea to compute the bootstrap estimates. When we bootstrap a quantile regression, we can use the sample estimates to guess the sign of the residuals in the bootstrap sample. With this algorithm, bootstrapping the quantile regression estimator is actually quicker than bootstrapping the OLS estimator. We cannot apply a  similar approach to least squares estimators because of  their global nature.

Instead of bootstrapping the quantile regression estimates, it is possible to bootstrap the score (or estimating equation) of the estimator. This approach amounts in fact to using the one-step estimator to compute the bootstrap estimate  when we take the sample estimate as a preliminary guess. This inference procedure, which has been suggested  for quantile regression by \cite{ChHa06} and \cite{belloni2017program}, is extremely  fast and can also be used to  perform uniform inference. Its drawback is the necessity to choose a smoothing parameter to estimate the conditional density of the response variable given the covariates.

The simulations we report in Section \ref{sec:simulations} show that the preprocessing algorithm is $30$ times faster than Stata's built-in algorithm when we have $50,000$ observations, $20$ regressors and we estimate $99$ quantile regressions. The one-step estimator further divides the computing time by almost $4$. The preprocessing step applied to the bootstrap  of a single quantile regression divides the computing time by about $10$. The score multiplier bootstrap further divides the computing time by $10$ compared to the preprocessing algorithm. Thus, these new algorithms open new possibilities for quantile regression methods.  For instance, in the application reported in Section \ref{sec:application}, we could estimate $91$ different quantile regressions in a sample of $2,194,021$ observations, with $14$ regressors and bootstrap $100$ times the estimates  in about 30 minutes on a laptop. The same estimation with the built-in commands of Stata would take over two months. The codes in Stata and some rudimentary codes in R are available from the authors.
 
We organize the rest of the paper as follows. Section \ref{sec:process} briefly defines the quantile regression model and estimator, describes the existing algorithms and provides the limiting distribution of the estimator. In Section \ref{sec:preprocessing}, we adapt the preprocessing step of \cite{portnoy1997gaussian} to estimate the whole quantile regression process. In Section \ref{sec:onestep}, we  suggest the new one-step estimator. Section \ref{sec:bootstrap} uses the same strategies to develop fast algorithms for bootstrap. Section \ref{sec:simulations} and \ref{sec:application} provide the results of the simulations and of the application, respectively. Finally, in Section \ref{sec:prospects}  we point out some directions of research that may be fruitful. 

\section{The quantile regression process}
\label{sec:process}

This section gives a very brief introduction to the linear quantile regression model. For a more thorough discussion we recommend the book written by \cite{Koe05} and the recent Handbook of Quantile Regression \citep{koenker2017handbook}.

\subsection{The quantile regression model}
\label{sec:model}
We are often interested in learning the effect of a $%
k\times 1$ vector of covariates (including a constant) $X$ on the distribution of a scalar response variable $Y$. Let $Q_Y(\tau|x )$ be the $\tau$ conditional quantile of $Y$ given $X=x$ with $0<\tau<1$. The 
conditional quantile function of $Y$ given $X$, $\tau \mapsto Q_{Y}\left(\tau | X \right) $%
, completely describes the dependence between $Y$ and $X$.
For computational convenience and simplicity of the interpretation, we assume that the  conditional quantile functions are linear in $x$:
\begin{assumption}[Linearity]
\label{assumption:linear}
\begin{equation*}
Q_{Y}\left( \tau |x\right) =x^{\prime }\beta \left( \tau \right)
\end{equation*}
for all $\tau \in \mathcal{T}$, which is a closed subset of $[\varepsilon,1-\varepsilon]$ for some $0<\varepsilon<1$. 
\end{assumption}

In this paper we focus on central quantile regression and suggest inference procedures justified by asymptotic normality arguments. Therefore, we must exclude the tail regions. For extremal quantile regressions \cite{chernozhukov2011inference} have suggested alternative inference procedures based on extreme value approximations. As a rule of thumb for continuous regressors, they suggest that normal inference performs satisfactorily if we set $\varepsilon=15\cdot k/n$.

Most existing sampling properties of the quantile regression estimator have been derived for continuous response variables. Beyond the technical reasons for this assumption, the linearity assumption for the conditional quantile functions is highly implausible for discrete or mixed discrete-continuous response variables. Therefore, we impose the following standard assumption:


\begin{assumption}[Continuity]
\label{assumption:continuity}
 The conditional density function $f_Y(y|x)$ exists, is uniformly continuous
in $(y, x)$ and bounded on the support of $(Y, X)$.
\end{assumption}

For  identification reasons, we require that the (density weighted) covariates are not linearly dependent:

\begin{assumption}[Rank condition]
\label{assumption:rank}
The minimal eigenvalue of the Jacobian matrix $J(\tau):=E[f_y(X'\beta(\tau)|X)XX')]$ is bounded away from zero uniformly over $\tau \in \mathcal{T}$.
\end{assumption}


Finally, we must impose a condition on the distribution of the covariates:

\begin{assumption}[Distribution of the covariates]
\label{assumption:covariates} \ 
$E\|X\|^{2+\varepsilon}<\infty$ for some $\varepsilon >0$.
\end{assumption}

Assumptions \ref{assumption:rank} and \ref{assumption:covariates} impose  that the derivatives of the coefficient function $\tau \mapsto \beta(\tau)$ are bounded uniformly on $\mathcal{T}$ because, by simple algebra (e.g., proof of Theorem 3 in \cite{Angrist+06}),
\begin{equation}\label{eq:deriv}
\frac{\mathrm{d} \beta(\tau)}{\mathrm{d} \tau} = J(\tau)^{-1} E(X).
\end{equation}


Under the linearity and the continuity assumptions, by the definition of the quantile function, the parameter $\beta(\tau)$ satisfies the following conditional moment restriction:
\begin{equation}
\label{eq:linear}
P(Y\leq x'\beta(\tau)|X=x)=\tau.
\end{equation}

It can be shown that $\beta(\tau)$ is the unique solution to the following optimization problem:
\begin{equation}
\label{eq:optimization}
\beta\left( \tau \right) =\underset{b\in\mathbb{R}^k}{\arg \min }E\left[ \rho _{\tau }\left( Y-X'b\right) \right]
\end{equation}
where $\rho_{\tau}$ is the check function defined as $\rho _{\tau }\left( U\right) =\left( \tau
-1\left( U\leq 0\right) \right) \cdot U$ and $1\left( \cdot \right) $ is the
indicator function.\footnote{See \cite{Koenker1978}.}
The objective function is globally convex and its first derivative with respect  to $b$ is $E\left[\left(\tau-1\left(Y\leq X'b\right)\right)X\right]$. Thus, $\beta(\tau)$ solves the unconditional moment condition
\begin{equation}
M\left(\tau,\beta\left(\tau\right)\right) := E\left[\left(\tau-1\left(Y\leq X'\beta\left(\tau\right)\right)\right)X\right]=0,
\end{equation}
which follows from the original conditional moment (\ref{eq:linear}).

\subsection{The quantile regression estimator}
\label{sec:estimator}

Let $\{y_i,x_i\}_{i=1}^n$ be a random sample from $\{Y,X\}$. The quantile regression estimator of \cite{Koenker1978} is the M-estimator that solves the sample analog of (\ref{eq:optimization}):
\begin{equation}
\label{eq:qr_estimator}
\widehat{\beta}\left( \tau \right) \in \underset{b\in \mathcal{R}^{k}}{\arg \min }
\sum_{i=1}^{n}\rho _{\tau }\left( y_{i}-x_{i}^{\prime }b\right) .
\end{equation}
This estimator is not an exact Z-estimator because the check function is not differentiable at $0$.  However, for continuous response variables, this will affect only the $k$ observations for which $y_i=x_i'\widehat\beta (\tau)$. Thus, the remainder term vanishes asymptotically and this estimator can be interpreted as an approximate Z-estimator: 
\begin{equation}
\label{eq:moment}
\widehat M\left( \tau ,\widehat{\beta}\left( \tau \right) \right) := \frac{1}{n}\sum_{i=1}^{n}\left(\tau -1\left(y_i\leq x_i'\widehat\beta\left(\tau\right)\right)\right)x_i=o_p\left(\frac{1}{\sqrt n}\right).
\end{equation}

The minimization problem (\ref{eq:qr_estimator}) that $\widehat{\beta}\left( \tau
\right) $ solves can be written as a convex linear program. This kind of
problem can be relatively efficiently solved with some modifications of the
simplex algorithm, see \cite{barrodale1974solution}, \cite{koenker1987algorithm} and \cite{koenker1994remark}. The modified simplex algorithm performs extremely well
for problems of moderate size but becomes relatively slow in larger samples. Worst-case
theoretical results indicate that the number of iterations required can
increase exponentially with the sample size. For this reason, \cite%
{portnoy1997gaussian} have developed an interior point algorithm.
Unlike the simplex, interior point algorithms start in the interior of
the feasible region of the linear programming program and travel on a path
towards the boundary, converging at the optimum. The inequality
constraints are replaced by a barrier that penalizes points that are close
to the boundary of the feasible set. Since this barrier idea was pioneered
by Ragnar Frisch and each iteration corresponds to a Newton step, \cite%
{portnoy1997gaussian} call their application of the interior point method to
quantile regression the Newton-Frisch algorithm. \cite%
{portnoy1997gaussian} have also suggested a preprocessing step that we will discuss in Section \ref{sec:preprocessing}.

\subsection{Sampling properties}
\label{sec:estimator}

For the sake of completeness we provide distribution theory for the quantile regression coefficient process. Following \cite{Angrist+06}, we do not impose Assumption \ref{assumption:linear}, therefore allowing for the possibility of model misspecification.

\begin{proposition}[Asymptotic distribution theory]
Under Assumptions 2 to 4, the quantile regression estimator defined in \eqref{eq:qr_estimator} is  consistent for the parameter defined in \eqref{eq:optimization} uniformly over $\mathcal{T}$. Moreover, the quantile regression process $J(\cdot)\sqrt n (\widehat\beta(\cdot)-\beta(\cdot))$ weakly converges to a zero mean Gaussian process $z(\cdot)$ in $\ell^{\infty}(\mathcal{T})^k$, where $\ell^{\infty}(\mathcal{T})$ is the set of bounded functions on $\mathcal{T}$ and $z(\cdot)$ is defined by its covariance function
\begin{equation*}
\Sigma(\tau,\tau')\equiv E[\left(\tau-1\left(Y\leq X'\beta(\tau)\right)\right)\left(\tau'-1\left(Y\leq X'\beta(\tau')\right)\right)XX'].
\end{equation*}
If the model is correctly specified, i.e. under Assumption 1, then $\Sigma(\tau,\tau')$ simplifies to 
\begin{equation*}
(\min(\tau,\tau')-\tau\tau')\cdot E[XX'].
\end{equation*}
\end{proposition}

Uniformly consistent estimators of $J(\tau)$ and  $\Sigma(\tau,\tau')$ are useful for analytical inference and will be required to implement the one-step estimator and the score bootstrap. We use the sample analog of $\Sigma(\tau,\tau')$ and \cite{powell1991estimation} kernel estimator of $J(\tau)$:
\begin{equation}
\label{acf}
\widehat{\Sigma}\left( \tau ,\tau'\right) =\frac{1}{n}\sum_{i=1}^{n}(\tau-1(y_i\leq x_i'\widehat\beta(\tau)))(\tau'-1(y_i\leq x_i'\widehat\beta(\tau')))x_ix_i'
\end{equation}
\begin{equation}
\label{eq:powell}
\widehat{J}\left( \tau \right) =\frac{1}{n\cdot h_{n}}\sum_{i=1}^{n}K\left( 
\frac{y_{i}-x_{i}\widehat{\beta}\left( \tau \right) }{h_{n}}\right)
x_{i}x_{i}^{\prime }
\end{equation}
where $K\left( \cdot \right) $ is a kernel function and $h_{n}$ is a
bandwidth. We use a standard normal density as kernel function and
the \cite{HallSheather} bandwidth%
\begin{equation}
h_{n}=n^{-1/3}\cdot \Phi ^{-1}\left( 1-\frac{\alpha }{2}\right) ^{2/3}\left[ 
\frac{1.5\cdot \phi \left( \Phi ^{-1}\left( \tau \right) \right) ^{2}}{%
2\cdot \Phi ^{-1}\left( \tau \right) ^{2}+1}\right] ^{1/3}
\label{Hall-Sheather}
\end{equation}
where $\phi \left( \cdot \right) $ and $\Phi ^{-1}\left( \cdot \right) $ are the density and quantile functions of the standard normal  and $\alpha $ is the targeted
level of the test.
These estimators are uniformly consistent under the additional assumption that $E\|X\|^{4}$ is finite. The pointwise variance can consistently be estimated by
\begin{equation}
\label{eq:point_se}
\widehat J(\tau)^{-1}\widehat\Sigma(\tau,\tau)\widehat J(\tau)^{-1}.
\end{equation}

\section{Preprocessing for the quantile regression process}
\label{sec:preprocessing}

\subsection{Portnoy and Koenker (1997)}
\label{sec:PK1997}

The fitted quantile regression line interpolates at least $k$ data points. It can easily be shown from the moment condition (\ref{eq:moment}) that the
quantile regression estimates are numerically identical if we change the
values of the observations that are not interpolated as long as they remain
on the same side of the regression line. The sign of the residuals $y_i-x_i'\widehat\beta(\tau)$ is the
only thing that matters in the determination of the estimates. This explains
why the quantile regression estimator is influenced only by the local
behavior of the conditional distribution of the response near the specified
quantile and is robust to outliers in the response variable if we do not go too far into the tails of the distribution.\footnote{On the other hand, note that quantile regression is not robust to outliers in the x-direction.}

\cite{portnoy1997gaussian} exploit this
property to design a quicker algorithm. Suppose for the moment that we knew that a
certain subset $J_{H}$ of the observations have positive residuals and
another subset $J_{L}$ have negative residuals. Then the solution to the
original problem (\ref{eq:qr_estimator}) is exactly the same as the solution to
the following revised problem%
\begin{equation}
\underset{b\in \mathcal{R}^{K}}{\min }\sum_{i\notin \left( J_{H}\cup
J_{L}\right) }\rho _{\tau }\left( y_{i}-x_{i}^{\prime }b\right) +\rho _{\tau
}\left( y_{L}-x_{L}^{\prime }b\right) +\rho _{\tau }\left(
y_{H}-x_{H}^{\prime }b\right)  \label{modified_problem}
\end{equation}%
where $x_{G}=\sum_{i\in J_{G}}x_{i}$, for $G\in \left\{ H,L\right\} $, and $%
y_{L},y_{H}$ are chosen small and large enough, respectively,
to ensure that the corresponding residuals remain negative and positive.
Solving this new problem gives numerically the same estimates but is
computationally cheaper because the effective sample size is reduced by the
number of observations in $J_{H}$ and $J_{L}$.

In order to implement this idea we need to have some preliminary information about
the sign of some residuals. \cite{portnoy1997gaussian} suggest to use only a subsample to estimate an initial quantile
regression that will be used to guess the sign of the residuals in the whole
sample. More formally, their algorithm works as follows:
\begin{algorithm}[Portnoy and Koenker (1997)]
\label{alg:pk1997}
\

\noindent

\begin{enumerate}
\item Solve the quantile regression problem (\ref{eq:qr_estimator}) using only
a subsample of size $\left( k\cdot n\right) ^{2/3}$ from the original
sample. This delivers a preliminary estimate $\tilde{\beta}\left( \tau
\right) .$

\item Calculate the residuals $r_{i}=y_{i}-x_{i}^{\prime }\tilde{\beta}%
\left( \tau \right) $ and $z_{i}$, a quickly computed conservative estimate
of the standard error of $r_{i}$. Calculate the $\tau -\frac{M}{2n}$ and $%
\tau +\frac{M}{2n}$ quantiles of $\frac{r_{i}}{z_{i}}$. The observations
below this first quantile are included in $J_{L}$; the observations above
this second quantile are included in $J_{H}$; the $M=m\cdot \left( k\cdot
n\right) ^{2/3}$ observations between these quantiles are kept for the next
step. $m$ is a parameter that can be chosen by the user; by default it is
set to $0.8$.

\item Solve the modified problem (\ref{modified_problem}) and obtain $\widehat{%
\beta}\left( \tau \right) $

\item Check the residual signs of the observations in $J_{L}$ and $J_{H}$:

\begin{enumerate}
\item If no bad signs (or the number of bad signs is below the number of
allowed mispredicted signs), $\widehat{\beta}\left( \tau \right) $ is the
solution.

\item If less than $0.1\cdot M$ bad signs: take the observations with
mispredicted sign out of $J_{L}$ and $J_{H}$ and go back to step 3.

\item If more than $0.1\cdot M$ bad signs: go back to step 1 with a doubled
subsample size.
\end{enumerate}
\end{enumerate}
\end{algorithm}

Formal computational complexity results in \cite{portnoy1997gaussian}
indicate that the number of computation required by the simplex is quadratic
in $n$, by the interior point method is of order $nk^{3}\log ^{2}n$, and by the
preprocessing algorithm is of order $n^{2/3}k^{3}\log ^{2}n+nk^{2}$.

\subsection{Preprocessing for the quantile regression process}
\label{sec:preprocess}

In many cases, the researcher is interested in the quantile regression coefficients at several quantiles. This allows analyzing the heterogeneity of behavior, which is the main motivation for using quantile regression. For instance, it is common to estimate $99$ quantile regressions, one at each percentile, and plot the estimated coefficients. Tests of functional hypotheses such as those suggested in \cite{Koenker2002}, \cite{ChFe05} and \cite{ChHa06} necessitate the estimation of a large number of quantile regressions. Some estimators also require the preliminary estimation of the whole quantile regression process, such as the estimators suggested in  \cite{koenker1987estimation}, \cite{MachadoMata05} and \cite{chernozhukov2013inference}.

In the population we can let the quantile index $\tau$ increase continuously from $0$ to $1$ to obtain a continuous quantile regression coefficient process. In finite samples, however, only a finite number of distinct quantile regressions exists. In the univariate case, there are obviously only $n$ different sample quantiles. In the general multivariate case, the number of distinct solutions depends on the specific sample but \cite{portnoy1991asymptotic} was able to show that the number of distinct quantile regressions is of order $O_p(n\log n)$ when the number of covariates is fixed. \cite{koenker1987algorithm} provides an efficient parametric linear programming algorithm that computes the whole quantile regression process.\footnote{Parametric programming is a technique for investigating the effects of a change in the parameters (here of the quantile index $\tau$) of the objective function. } Given $\widehat\beta(\tau)$, a single simplex pivot is required to obtain the next quantile regression estimates. This is a feasible solution when $n$ and $k$ are not too large but it is not feasible to compute  $O_p(n\log n)$ different quantile regressions in the type of applications that we want to cover. Therefore, we do not try to estimate the whole quantile regression process. Instead, we discretize the quantile regression process and estimate quantile regressions only on a grid of quantile indexes such as $\tau=0.01,0.02,...,0.98,0.99$.

In order to approximate well enough the conditional distribution, several estimators requires that the mesh width of the grid converges to $0$ at the $n^{-1/2}$ rate or faster.\footnote{See e.g.  \cite{chernozhukov2013inference}.} \cite{neocleous2008monotonicity} show that it is enough to estimate quantile regressions on a grid with a mesh width of order $n^{-1/4}$ if we linearly interpolates between the estimated conditional quantiles and an additional smoothness condition is satisfied (the derivative of the conditional density must be bounded). Thus, interpolation may be used in a second step to improve the estimated conditional quantile function but even in this case  a significant number of quantile regressions must be estimated.

At the moment, when a researcher wants to estimate 99 quantile regressions, she must estimate them separately and the
computation time will be 99 times longer than the time needed for a single
quantile regression.\footnote{It is actually possible to use the estimates from the previous quantile regression as starting values for the next quantile regression. These better starting values allow for reducing the computing time and are, therefore, used by all our algorithms.} We build on the preprocessing idea of \cite%
{portnoy1997gaussian} and suggest an algorithm that exploits recursively the
quantile regressions that have already been estimated in order to estimate
the next one.\footnote{%
In his comment of \cite{portnoy1997gaussian}, \cite{thisted1997gaussian}
suggests this idea, which has never been implemented to the best of our knowledge.%
}

Assume we want to estimate $J$ different quantile regressions at the
quantile indexes $\tau _{1}<\tau _{2}<...<\tau _{J}$.  

\begin{algorithm}[Preprocessing for the quantile regression process]
\label{alg:prepro}
\

\noindent
To initialize
the algorithm, $\widehat{\beta}\left( \tau _{1}\right) $ is estimated using one of the traditional algorithms described above.
Then, iteratively for $j=2,...,J$:
\begin{enumerate}
\item Use $\widehat\beta\left( \tau_{j-1}\right)$ as a preliminary estimate.

\item Calculate the residuals $r_{i}=y_{i}-x_{i}^{\prime }\widehat{\beta}%
\left( \tau_{j-1} \right) $ and $z_{i}$, a quickly computed conservative estimate
of the standard error of $r_{i}$. Calculate the $\tau -\frac{M}{2n}$ and $%
\tau +\frac{M}{2n}$ quantiles of $\frac{r_{i}}{z_{i}}$. The observations
below this first quantile are included in $J_{L}$; the observations above
this second quantile are included in $J_{H}$; the $M=m\cdot \left( k\cdot
n\right) ^{1/2}$ observations between these quantiles are kept for the next
step. $m$ is a parameter that can be chosen by the user; by default it is
set to $3$.

\item Solve the modified problem (\ref{modified_problem}) and obtain $\widehat{%
\beta}\left( \tau_j \right) $

\item Check the residual signs of the observations in $J_{L}$ and $J_{H}$:

\begin{enumerate}
\item If no bad signs (or the number of bad signs is below the number of
allowed mispredicted signs), $\widehat{\beta}\left( \tau_j \right) $ is the
solution.

\item If less than $0.1\cdot M$ bad signs: take the observations with
mispredicted sign out of $J_{L}$ and $J_{H}$ and go back to step 3.

\item If more than $0.1\cdot M$ bad signs: go back to step 2 with a doubled
$m$.
\end{enumerate}
\end{enumerate}
\end{algorithm}

Compared to Algorithm \ref{alg:pk1997}, the preliminary estimate does not need to be computed because we can take the already computed $\widehat\beta(\tau_{j-1})$. This will naturally provide a good guess of the sign of the residuals only if $\tau_j$ and $\tau_{j-1}$ are close.  This can be formalized by assuming that 
 $\sqrt n(\tau_j-\tau_{j-1})=O_p\left(1\right)$. When this condition is satisfied, then $\sqrt n(\widehat\beta(\tau_{j})-\widehat\beta(\tau_{j-1}))=O_p(1)$ by the stochastic equicontinuity of the quantile regression coefficient process. This justifies keeping a smaller sample in step 2: while we kept a sample proportional to $n^{2/3}$ in Algorithm \ref{alg:pk1997}, we can keep a sample proportional to $n^{1/2}$ in Algorithm \ref{alg:prepro}. 

These two differences (no need for a preliminary estimate and smaller sample size in the final regression) both imply that Algorithm \ref{alg:prepro} is faster than Algorithm \ref{alg:pk1997} and should be preferred when a large number of quantile regressions is estimated. Finally, note that  step 4 of both algorithms makes sure that the estimates are numerically equal (or very close if we allow for a few mispredicted signs of the residuals) to the estimates that we would obtain using  the  simplex or interior point algorithms. Thus, there is no statistical trade-off to consider when deciding which algorithm should be used. This decision can be based purely on the computing time of the algorithms. In the next section, we will consider an even faster algorithm but it will not provide numerically identical estimates. This new estimator will be only asymptotically equivalent to the traditional quantile regression estimator that solves the optimization problem (\ref{eq:qr_estimator}).

\section{One-step estimator}
\label{sec:onestep}

One-step estimators were introduced for their asymptotic efficiency by \cite{le1956asymptotic}. A textbook treatment of this topic can be found in Section 5.7 of 
\cite{vanderVaart98}. The one-step method builds on and improves a preliminary estimator. If this preliminary estimator is already $\sqrt n$-consistent, then the estimator that solves a linear approximation to the estimating equation is asymptotically equivalent to the estimator that solves the original estimating equation. In other words, when we start from a  $\sqrt n$-consistent starting value, then a single Newton-Raphson iteration is enough. Further iterations do not improve the first-order asymptotic distribution.

In this section we suggest a new estimator for the quantile regression coefficient process. This estimator is asymptotically equivalent to the estimators presented in the two preceding sections but may differ in finite samples. The idea consists in starting from one quantile regression computed using a standard algorithm and then obtaining sequentially the remaining regression coefficients using a single Newton-Raphson iteration for each quantile regression.\footnote{We start from the median regression in the simulations and application in Sections \ref{sec:simulations} and \ref{sec:application}.} In other words, this is a one-step estimator that uses a previously estimated quantile regression for a similar quantile $\tau$ as the preliminary estimator.

Assume that we have already obtained $\widehat{\beta}\left(
\tau_1\right) $, a $\sqrt n$-consistent estimator of $\beta(\tau_1)$, and we
would like to obtain $\widehat{\beta}\left( \tau_1+\varepsilon \right) $ for a small $%
\varepsilon $. Formally, we assume that  $\varepsilon = O(n^{-1/2})$ such that $\widehat{\beta}\left(
\tau_1\right) $ is $\sqrt n$-consistent for $\beta(\tau_1+\varepsilon)$ because, by the triangle inequality,
$$
\| \widehat{\beta}\left(\tau_1\right) - \beta(\tau_1+\varepsilon)\| \leq \| \widehat{\beta}\left(\tau_1\right) - \beta(\tau_1)\| + \sup_{\tau_1 \leq \tau' \leq \tau_1+\varepsilon} \left \| \frac{\mathrm{d} \beta(\tau')}{\mathrm{d} \tau} \right\| \varepsilon = O_P(n^{-1/2}) 
$$
where the last equality follows from \eqref{eq:deriv} together with Assumptions \ref{assumption:rank} and \ref{assumption:covariates}. Then, by Theorem 5.45 in \cite{vanderVaart98} the one step estimator
$$
\widehat{\beta}\left( \tau_1+\varepsilon \right) =\widehat{\beta}\left( \tau_1\right) -\widehat J(\tau_1) ^{-1}M\left( \tau_1+\varepsilon ,\widehat{\beta}\left( \tau_1\right) \right)
$$ 
has the same first-order asymptotic distribution as the quantile regression  estimator of $\beta(\tau_1 + \varepsilon)$. Here  we use that 
$$
\| \widehat J(\tau_1)  - J(\tau_1 + \varepsilon) \| \leq \| \widehat J(\tau_1)  - J(\tau_1 ) \| + \|J(\tau_1 )  - J(\tau_1 + \varepsilon)  \| \to_P 0,
$$
by the triangle inequality, Assumptions \ref{assumption:continuity} and \ref{assumption:covariates},  standard regularity conditions for $\widehat J(\tau_1)  \to_P J(\tau_1 )$, and $\varepsilon \to 0$. Note that the previous argument holds for any $\tau_1,\tau_1 + \varepsilon \in \mathcal{T}$ such that $\varepsilon= O(n^{-1/2})$.

The one-step estimator corresponds to a single Newton-Raphson iteration. It is possible
to use the resulting value as the new starting value for a second iteration and so on, but the first-order asymptotic distribution does not change. If we iterate until convergence, then $$J(\tau_1) ^{-1}M\left( \tau_1+\varepsilon ,\widehat{\beta}\left( \tau_1+\varepsilon\right) \right)=0$$Since $\widehat J(\tau_1)$ is asymptotically full rank by Assumption \ref{assumption:rank}, this implies that the moment condition (\ref{eq:moment}) must be satisfied at $\tau_1+\varepsilon$ such that we obtain numerically the same values as the traditional quantile regression estimator of $\beta(\tau_1+\varepsilon)$. This property, together with the fact that the quantile regression estimate is constant over a small range of quantile indexes, also implies that we can get numerically identical estimates to the traditional quantile regression estimator by choosing $\varepsilon$ to be small enough. This is, however, not the objective because the computation time of such a procedure would necessarily be higher than that of the parametric linear programming algorithm.\footnote{\cite{schmidt2016quantile} have suggested a different iterative estimation strategy. They also start from one quantile regression but they add or subtract sums of nonnegative functions to
it to calculate other quantiles. Their procedure has a different objective (monotonicity of the estimated conditional quantile function) and their estimator is not asymptotically equivalent to the traditional quantile regression estimator.}


The algorithm is summarized below:

\begin{algorithm}[One-step estimator for the quantile regression process]
\label{alg:onestep}
\

\noindent
To initialize
the algorithm, $\widehat{\beta}\left( \tau _{1}\right) $ is estimated using one
of the traditional algorithms described above.
Then, iteratively for $j=2,...,J$:
\begin{enumerate}
\item Use $\widehat\beta\left( \tau_{j-1}\right)$ as a preliminary estimate.

\item Estimate the Jacobian matrix with \cite{powell1991estimation} estimator and \cite{HallSheather} bandwidth and obtain $\widehat J(\tau_{j-1})$. 

\item Update the quantile regression coefficient \begin{equation}
\label{eq:onestep}
\widehat{\beta}\left( \tau_j \right) 
=\widehat{\beta}\left( \tau_{j-1} \right) -\widehat{J}\left( \tau_{j-1} \right) ^{-1}\frac{1}{n}\sum_{i=1}^{n}\left( \tau_j -1\left( y_{i}\leq x_{i}\widehat{\beta}\left( \tau_{j-1} \right) \right)
\right)
\end{equation}

\end{enumerate}
\end{algorithm}


The one-step estimator is much faster to compute when we do not choose a too fine grid of quantiles. Our simulations reported in Section \ref{sec:simulations} show that a grid with $\varepsilon=0.01$ works well even for large samples. This result depends naturally on the data generating process. A fine grid may be needed when the conditional density of $Y$  is  changing quickly (i.e. when the first order approximation error is large). In practice, it is possible to estimate a few quantile
regressions with the traditional estimator and check that the difference between both estimators is small.

\section{Fast algorithms for bootstrap inference}
\label{sec:bootstrap}

Researchers have  often found  that the  bootstrap is the most reliable method for both pointwise and uniform inference. Naturally, this method is also the most demanding computationally. The time needed to compute the estimates can be a binding constraint when researchers are considering bootstrapping strategies. Fortunately, the same approaches that we have presented in the preceding sections (preprocessing and linear approximations) can also fruitfully reduce the computing time of the bootstrap.

\subsection{Preprocessing for the bootstrap}
\label{sec:preboot}

A very simple modification of the preprocessing algorithm of \cite{portnoy1997gaussian} leads
to significant improvements. The advantage when we compute a quantile
regression for a bootstrap sample is that we can use the already computed
estimate in the whole sample to guess the sign of the residuals. This means that
we can skip step 1 of the preprocessing Algorithm \ref{alg:pk1997}. In addition, this preliminary estimate is more precise
because it was computed using a sample of size $n$ instead of $\left( k\cdot
n\right) ^{2/3}$ in the original preprocessing algorithm. Thus, we need only to keep a lower
number of observations in step 2. We choose $M=m\cdot \left( k\cdot
n\right) ^{1/2}$ where $m$ is set by default to $3$ but can be modified by
the user. This multiplying constant $3$ was chosen because it worked well in our simulations. We do not have a theoretical justification for this choice and further improvements should be possible. In particular, it should be possible to adjust this constant during the process (that is after the estimates in a few bootstrap samples have been computed) by increasing it when the sign of too many residuals is mispredicted in step 4 or decreasing it when the sign of the residuals is almost never mispredicted.\footnote{A similar idea could be applied to adjust the constant $m$ in Algorithm \ref{alg:prepro}. The additional difficulty is that the optimal constant probably depends on the quantile index $\tau$, which is not the case for the bootstrap.}

We now define formally the algorithm for the empirical bootstrap but a similar algorithm can be applied to all types of the exchangeable bootstrap.

\begin{algorithm}[Preprocessing for the bootstrap]
\label{alg:preboot}
\

\noindent
For each bootstrap iteration $b=1,...,B$, denote by $\{y_i^{*b},x_i^{*b}\}_{i=1}^n$ the bootstrap sample:
\begin{enumerate}
\item Use $\widehat\beta\left( \tau \right)$ as a preliminary estimate.

\item Calculate the residuals $r_{i}^{*b}=y_{i}^{*b}-x_{i}^{*b\prime }\widehat{\beta}%
\left( \tau \right) $ and $z_{i}^{*b}$, a quickly computed conservative estimate
of the standard error of $r_{i}^{*b}$. Calculate the $\tau -\frac{M}{2n}$ and $%
\tau +\frac{M}{2n}$ quantiles of $\frac{r_{i}^{*b}}{z_{i}^{*b}}$. The observations
below this first quantile are included in $J_{L}$; the observations above
this second quantile are included in $J_{H}$; the $M=m\cdot \left( k\cdot
n\right) ^{1/2}$ observations between these quantiles are kept for the next
step. $m$ is a parameter that can be chosen by the user; by default it is
set to $3$.

\item Solve the modified problem (\ref{modified_problem}) for the sample $\{y_i^{*b},x_i^{*b}\}_{i=1}^n$and obtain $\widehat{\beta}^{*b}\left( \tau \right)$.

\item Check the residual signs of the observations in $J_{L}$ and $J_{H}$:

\begin{enumerate}
\item If no bad signs (or the number of bad signs is below the number of
allowed mispredicted signs), $\widehat{\beta}^{*b}\left( \tau \right)$, is the
solution.

\item If less than $0.1\cdot M$ bad signs: take the observations with
mispredicted sign out of $J_{L}$ and $J_{H}$ and go back to step 3.

\item If more than $0.1\cdot M$ bad signs: go back to step 2 with a doubled
$m$.
\end{enumerate}
\end{enumerate}
\end{algorithm}

With this algorithm, bootstrapping a single quantile regression becomes \emph{faster} than bootstrapping the least squares estimator. The preprocessing strategy does not apply to least squares estimators because of the global nature of these estimators.

When the whole quantile regression process is bootstrapped, either Algorithm \ref{alg:prepro}  or Algorithm \ref{alg:preboot} can be applied.\footnote{Algorithm \ref{alg:prepro} can be slightly improved by using preprocessing with   $\hat\beta(\tau_1)$ as a preliminary estimate of    $\hat\beta^{*b}(\tau_1)$ instead of computing it completely from scratch.  } In our simulations we found that the computing times were similar for these two algorithms. In our implementation, Algorithm   \ref{alg:preboot} is used in this case. Even shorter computing times
can be obtained either by using the one-step estimator in each bootstrap sample or by bootstrapping the linear representation of the estimator, which we will present in the next subsection. These two algorithms are
not numerically identical to the traditional quantile regression estimator and require the choice of a smoothing parameter to estimate the Jacobian matrix.

\subsection{Score resampling (or one-step bootstrap)}
\label{sec:score_boot}
Even with the preprocessing of Algorithm \ref{alg:preboot}, we must recompute the
estimates in each bootstrap draw, which is naturally
computationally demanding. Instead of resampling the covariates and the response to
compute the coefficients, we can resample the asymptotic linear (Bahadur) representation of the estimators. \cite{belloni2017program} suggest and prove the validity of the multiplier score bootstrap. For $b=1,..., B$ we obtain the corresponding bootstrap draw of $\widehat{\beta}\left( \tau\right) $ via%
\begin{equation}
\label{eq:score_boot}
\widehat{\beta}^{\ast b}\left( \tau\right) =\widehat\beta(\tau)-\widehat{J}\left( u\right) ^{-1}\frac{1}{n}%
\sum_{i=1}^{n}\xi_{i}^{*b}\left( \tau -1\left( y_i\leq x_i'\widehat{\beta}\left( \tau \right) \right)
\right)
\end{equation}
where $\left(\xi _{i}^{*b}\right) _{i=1}^{n}$ are iid random variables that are independent
from the data and must satisfy the following restrictions:%
\begin{equation*}
E\left[ \xi _{i}^{*b}\right] =0\text{ and }Var\left( \xi _{i}^{xb}\right) =1.
\end{equation*}%
We have implemented this score multiplier bootstrap for three different
distributions of $\xi _{i}^{*b}$ (i) $\xi _{i}^{*b}\sim \mathcal{E}-1$ where $%
\mathcal{E}$ is a standard exponential random variable, which corresponds to
the Bayesian bootstrap (see e.g. \cite{hahn1997bayesian}), (ii) $\xi _{i}^{*b}\sim N\left( 0,1\right) $, which
corresponds to the Gaussian multiplier bootstrap (see e.g. \cite{gine1984some}), (iii) $\xi _{i}^{*b}\sim N_{1}/\sqrt{2}+\left( N_{2}^{2}-1\right) /2$ where $N_{1} $ and $N_{2}$ are mutually independent standard normal random
variables, which corresponds to the wild bootstrap (see e.g. \cite{mammen1993bootstrap}). In addition to the multiplier bootstrap, we have also implemented the score bootstrap suggested in \cite{ChHa06}, which corresponds to (\ref{eq:score_boot}) with multinomial weights.
Since these different distributions give very similar results, we report only the performance of the wild score bootstrap in the simulations in Section \ref{sec:simulations}.

The score resampling procedure can be interpreted as a one-step estimator of the bootstrap value where we use the sample estimate as the preliminary value. This interpretation can be seen by comparing equation (\ref{eq:onestep}) with equation (\ref{eq:score_boot}). \cite{kline2012score} notice this connection between the score bootstrap and one-step estimators in their remark 3.1. 
  
While the score bootstrap is much faster to compute than the
exchangeable bootstrap, it requires the preliminary estimation of the matrix 
$J\left( \tau\right) $. This matrix is a function of $f_{Y}\left( X'\beta(\tau)|X\right) $, which is relatively difficult to estimate and
necessitates the choice of a smoothing parameter. Thus, the multiplier
bootstrap has no advantage in this respect compared to analytical
estimators of the pointwise standard errors. On the other hand, the score bootstrap can be used to test functional hypotheses within relatively limited computing time (see \cite{ChHa06} for many examples).

To summarize, for bootstrapping the quantile regression process we can recompute the estimate in each bootstrap draw and each quantile using preprocessing as introduced in Section \ref{sec:preboot}. Alternatively, we can use a first-order approximation starting from another quantile regression in the same bootstrap sample (one-step estimator) or we can use a first-order approximation starting from the sample estimate at the same quantile (score resampling). The simulations in the following section will compare these different approaches both in term of computing time as in term of the quality of the inference. 

\section{Simulations}
\label{sec:simulations}

The results in Sections \ref{sec:simulations} and  \ref{sec:application} were obtained using Stata 14.2. We compare the official  Stata command (qreg), which implements a version of the simplex algorithm according to the documentation, with our own implementation of the other algorithms directly in Stata. The codes and the data to replicate these results are available from the authors.
The reference machine used for all the simulations is an
AMD Ryzen Threadripper 1950X with 16 cores at 3.4 GHz and 32 GB of
RAM (but for each simulation each algorithm exploits only
one processor, without any parallelism).

\subsection{Computation times}
\label{MCalgorithms}

We use empirical data to simulate realistic samples. We calibrate the data generating processes to match many characteristics of a Mincerian wage regression.  In particular, we draw covariates (such as education in years and as a set of indicator variables, a polynomial in experience, regional indicator variables, etc.) from the Current Population Survey data set from 2013. We consider different sets of regressors with dimensions ranging from 8 to 76. Then, we simulate the response variable by drawing observations from the estimated conditional log wage distribution. 

\begin{table}
\begin{center}
\captionsetup{font=large}
\caption{Comparison of the algorithms}
\label{table:algorithms}
\smallskip
\begin{tabular}{lccccccc}
\hline \hline \noalign{\smallskip} \multicolumn{8}{c}{Number of observations}\\
Algorithm & 500 & 1000 & 5000 & 10k & 50k & 100k & 500k\\
\noalign{\smallskip}\hline \hline \noalign{\smallskip}
\multicolumn{8}{l}{Panel 1: Estimation of a single quantile regression}\\
\multicolumn{8}{l}{A. Computing time in seconds for $\tau=0.5$ and $k=8$}\\
qreg (simplex) & 0.05 & 0.06 & 0.13 & 0.24 & 1.35 & 2.92 & 16.84\\
fn & 0.02 & 0.03 & 0.12 & 0.21 & 0.95 & 1.98 & 13.04\\
pqreg & 0.02 & 0.02 & 0.07 & 0.12 & 0.62 & 1.23 & 6.36\\
pfn  & 0.04 & 0.06 & 0.10 & 0.16 & 0.61 & 1.08 & 5.08\\
\multicolumn{8}{l}{B. Computing time in seconds for $\tau=0.5$ and $k=76$} \\
qreg  (simplex) &  0.26 & 0.43 & 2.14 & 4.80 & 37.75 & 85.69 & 622.4\\
fn &  0.06 & 0.08 & 0.25 & 0.41 & 1.93 & 3.72 & 20.78\\
pqreg & 0.17 & 0.23 & 0.80 & 2.51 & 11.36 & 22.08 & 89.44\\
pfn & 0.09 & 0.11 & 0.30 & 0.73 & 2.47 & 4.30 & 17.61\\
\noalign{\smallskip}\hline \noalign{\smallskip}
\multicolumn{8}{l}{Panel 2: Estimation of the whole QR process (99 quantile regressions, $k=20$)}\\
\multicolumn{8}{l}{Computing time in seconds} \\
qreg (simplex) & 5.93 & 9.01 & 48.86 & 119 & 1,084 &  & \\
preprocessing, qreg & 0.64 & 0.85 & 2.93 & 6.96 & 57.70 &  & \\
preprocessing, fn & 2.02 & 1.90 & 3.95 & 6.55 & 30.67 &  & \\
one-step estimator & 0.40 & 0.45 & 1.00 & 1.77 & 8.29 &  & \\
\noalign{\smallskip}\hline \noalign{\smallskip}
\multicolumn{8}{l}{Panel 3: Bootstrap of the median regression ($50$ bootstrap replications, $k=20$)}\\
\multicolumn{8}{l}{Computing time in seconds} \\
 bsqreg  (simplex) & 2.63 & 3.14 & 8.19 & 15.06 & 99.89 & 163.3 & \\
preprocessing & 0.42 & 0.63 & 1.89 & 3.50 & 13.55 & 18.52 & \\
multiplier & 0.08 & 0.10 & 0.25 & 0.60 & 1.36 & 1.90 & \\
\noalign{\smallskip}\hline\hline\end{tabular}\\
\end{center}
\small \qquad Note: Average computing times over $200$ replications.
\end{table}

Table \ref{table:algorithms} provides the average computing times (over 200 replications)  for three different cases. In the first panel, we compare different algorithms to estimate a single quantile regression.\footnote{We provide the results for the median regression but the ranking was similar at other quantile indexes.} It appears that the interior point algorithm (denoted by fn in the table) is always faster than the simplex algorithm as implemented by Stata. Preprocessing (denoted by pfn and pqreg) becomes an attractive alternative
when the number of observations is `large enough', where `large enough' corresponds to $5,000$ observations when there are $8$ regressors but $500,000$ observations when there are $76$ regressors. In this last configuration ($n=500,000$ and $k=76$) the computing time needed by the interior point algorithm with the preprocessing step is $35$ times lower than the computing time of the built-in Stata's command. The improvement in this first panel is only due to the implementation of algorithms suggested by \cite{portnoy1997gaussian}. All these estimators provide numerically identical estimates.

The second panel of Table \ref{table:algorithms} compares the algorithms
when the objective is to estimate 99 different quantile regressions, one at each percentile. For all sample sizes considered the algorithm with the preprocessing step suggested in Section \ref{sec:preprocess} computes the same estimates at a fraction of the computing time of the official Stata command. When $n=50,000$, the computing time is divided by more than $50$. The one-step estimator defined in Section \ref{sec:onestep} further divides the computing time by a factor of about $4$. Since this estimator is only asymptotically equivalent to the other algorithms, we also measure its performance compared to the traditional quantile regression estimator in Subsection \ref{sec:perf_one}.

In the third panel of Table \ref{table:algorithms} we compare the computing time needed to bootstrap $50$ times the median regression with the official Stata command, with the preprocessing step introduced in Section \ref{sec:preboot} and with the score bootstrap described in Section \ref{sec:score_boot}. The results show that the preprocessing step divides the computing time by about 4 for small sample sizes and by 9 for the largest sample size that we have considered. Using the multiplier bootstrap divides the computing time one more time by a factor of $10$. However, this type of bootstrap is not numerically identical to the previous ones. The simulations in the subsections \ref{sec:pointwise} and \ref{sec:functional} show that it tends to perform slightly worse in small samples.

\subsection{Performance of the one-step estimator}
\label{sec:perf_one}

\begin{figure}
\begin{center}
\includegraphics[width=12cm]{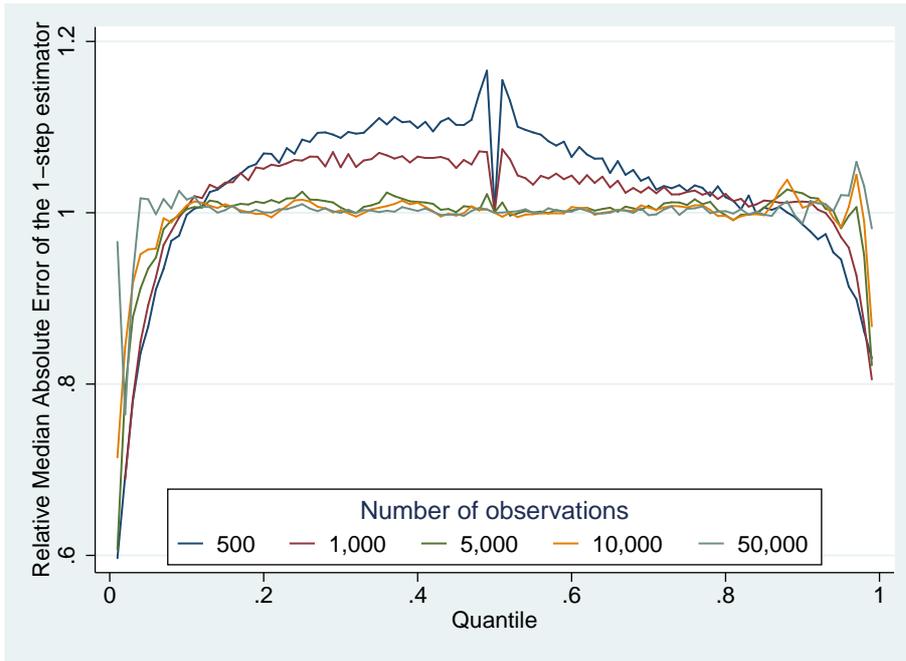}
\caption{Relative MAE as a function of the quantile and the number of observations}
\label{fig:perf_one}
\end{center}
\end{figure}

Table \ref{table:perf_one} and Figure \ref{fig:perf_one} measure the performance of the one-step estimator. We use the same data generating process as in the previous subsection with $k=20$ regressors. We perform $10,000$ replications to get precise results. We first note that   the one-step algorithm sometimes does not converge.
The reason is that $\widehat J(\tau)$  
is occasionally singular or close to singular in small samples. As a consequence, the one-step estimate  $\widehat \beta (\tau+\varepsilon)$, which is linear in $\widehat J(\tau)^{-1}$, takes a very large value. Once the estimated quantile regression coefficients at one quantile index are very far away from the true values, then the Jacobian   $\widehat J(\tau)$  
and the coefficients at the remaining quantile indices will be very far from the true values. It would be possible to detect convergence issues by checking that the moment conditions are approximately satisfied or by estimating a few quantile regressions with a traditional algorithm and checking that the estimates are close. We did not yet implement this possibility because the  convergence problem appears only in small samples (when $n\leq1000$) and traditional algorithms are anyway fast enough for these sample sizes.

Our parameters of interest are the $20\times 1$ vectors of quantile regression coefficients at the quantile indices $\tau=0.01,0.02,...0.99$. Thus, in principle, there are $1980$ different parameters of interest and we could analyze each of them separately.  Instead, we summarize the results and report averages  (over quantile indices and regressors) of several measures of performance    in Table \ref{table:perf_one}.\footnote{To make the estimates comparable across quantiles and regressors, we first normalize them such that they have unit variance in the specification with $n=50,000$. Then, we calculate the measures of performance separately for each parameter and average them over all quantile indices and regressors. The reported relative MSE and MAE are the averaged relative MSE and MAE. Alternatively, it is possible to calculate the ratio of the averaged MSE and MAE with the results in Table \ref{table:perf_one}. These ratios of averages and averages of ratios are very similar.} In Figure \ref{fig:perf_one} we plot the main measure separately for each quantile (but averaged over the regressors since the results are similar for different regressors).
The second part of Table \ref{table:perf_one} reports the standard measures of the performance of an estimator: squared (mean) bias, variance, and mean squared error (MSE). Since these measures based on squared errors are very sensitive to outliers, in the third part of the table, we also report measures based on absolute deviations: the absolute median bias, median absolute deviation (MAD) around the median, and median absolute error (MAE) around the true value.

\begin{table}
\begin{center}
\captionsetup{font=large}
\caption{Performance of the one-step estimator}
\label{table:perf_one}
\smallskip
\begin{tabular}{lcccccc}
\hline \hline \noalign{\smallskip} Number of obs. & 500 & 1000 & 5000 & 10k & 50k & 50k\\
Quantile step $\varepsilon$ for the 1-step est. & 0.01 & 0.01 & 0.01 & 0.01 & 0.01 & 0.001\\
\noalign{\smallskip}\hline \hline \noalign{\smallskip}
\multicolumn{7}{l}{Convergence of the one-step estimator:}\\
proportion converged & 0.581 & 0.754 & 0.999 & 1.000 & 1.000 & 1.000\\
\noalign{\smallskip}\hline \noalign{\smallskip}
\multicolumn{7}{l}{Measures based on squared errors (averages over 99 quantiles and 20 coefficients):}\smallskip\\
squared bias, QR & 0.383 & 0.174 & 0.055 & 0.037 & 0.020 & \\
squared bias, 1-step & 0.364 & 0.178 & 0.070 & 0.054 & 0.040 & 0.019
\smallskip\\
variance, QR & 141.6 & 62.68 & 10.70 & 5.182 & 1.000 & \\
variance, 1-step & 156.9 & 66.99 & 10.76 & 5.119 & 0.996 & 0.991
\smallskip\\
MSE, QR & 142.0 & 62.85 & 10.76 & 5.219 & 1.020 & \\
MSE, 1-step & 157.2 & 67.17 & 10.83 & 5.173 & 1.035 & 1.010\\
relative MSE of 1-step & 1.087 & 1.052 & 1.001 & 0.988 & 1.015 & 0.997 \\
\noalign{\smallskip}\hline \noalign{\smallskip}
\multicolumn{7}{l}{Measures based on absolute errors (averages over 99 quantiles and 20 coefficients):}\smallskip\\
absolute median bias, QR & 0.416 & 0.281 & 0.166 & 0.138 & 0.100 & \\
absolute median bias, 1-step & 0.476 & 0.317 & 0.201 & 0.175 & 0.151 & 0.100
\smallskip \\
MAD around the median, QR & 7.719 & 5.190 & 2.173 & 1.518 & 0.666 & \\
MAD around the median, 1-step & 8.004 & 5.290 & 2.161 & 1.507 & 0.669 & 0.665
\smallskip \\
MAE, QR & 7.731 & 5.197 & 2.177 & 1.523 & 0.672 & \\
MAE, 1-step & 8.021 & 5.303 & 2.172 & 1.518 & 0.685 & 0.671\\
relative MAE of 1-step & 1.030 & 1.015 & 0.995 & 0.996 & 1.019 & 1.001\\
\noalign{\smallskip}\hline\hline\end{tabular}\\
\end{center}
\small
\quad Note: Statistics computed over $10,000$ replications. 
\end{table}

The bias of the one-step estimator is slightly larger than the bias of the traditional estimator in small and moderate sample sizes but it contributes only marginally to the MSE and MAE because it is small. When $n=500$, the MAE (MSE) of the one-step estimator is on average $3\%$ ($8.7\%$) higher than the MAE (MSE) of the traditional quantile regression estimator. These  disadvantages  converge quickly to $0$ when the sample size increases and are negligible when we have at least $5,000$ observations. Thus, the one-step estimator performs well exactly when we need it, i.e. when the sample size is large and computing the traditional quantile regression estimator may take too much time.

In the last column of Table \ref{table:perf_one}, we can assess the role of $\varepsilon$, the distance between quantile indices. In the first $6$ columns, we use $\varepsilon=0.01$ and estimate $99$ quantile regressions for $\tau=0.01,0.02,...,0.99$. To satisfy the conditions required to prove the asymptotic equivalence of the one-step and the traditional estimators, we should decrease $\varepsilon$ when the sample size increases. The fixed difference between quantile indices does not seem to prevent the relative MAE and MSE of the one-step estimator to converge to $1$ when $n=10,000$. However, when $n=50,000$ we see a (slight) increase in the relative MSE and MAE of the one-step estimator. Therefore, in the last column of the table, we reduce the quantile step to $\varepsilon=0.001$ and we see that the relative MAE and MSE go back to $1$ as predicted by the theory. Of course, the computing time increases when the quantile step decreases such that the researcher may prefer to pay a (small) price in term of efficiency to compute the estimates more quickly.

Figure \ref{fig:perf_one} plots the relative MAE of the one-step estimator as a function of the quantile index. Remember that we initialize the algorithm at the median such that both estimators are numerically identical at that quantile. With $500$ or $1,000$ observations, we nevertheless observe the largest relative MAE at quantile indices close to the median. On the other hand, the one-step estimator is more efficient than the traditional estimator at the tails of the distribution. Given that the extreme quantile regressions are notoriously hard to estimate, we speculate that the one-step estimator may be able to reduce the estimation error by using implicitly some information from less extreme quantile regressions, similar to the extrapolation estimators for extreme quantile regression of \cite{wang:2012} and \cite{he:2016}. When the sample size increases, the curve becomes flatter and the relative MAE is very close to $1$ at all quantiles between the first and ninth decile. 

\subsection{Pointwise inference}
\label{sec:pointwise}

There exist several inference methods for quantile regression with very different computational burdens. In this subsection we compare  the methods for pointwise inference and the next subsection the methods for uniform inference. In both cases we simulate data from the same data generating process as \cite{hagemann2017cluster}:
\begin{equation}
\label{eq:dgp}
y_i=x_i+(0.1+x_i^2)\cdot u_i
\end{equation}
where $x_i\sim N(0,1)$ and $u_i\sim N(0,1/3)$. Table \ref{table:pointwise} provides the empirical rejection probabilities of a correct and of an incorrect null hypothesis about the coefficient on $x_i^2$ at a single quantile index. The intended statistical level of the tests is 5\%. We consider separately two different quantile indexes $\tau=0.5$ and $\tau=0.85$.   

We compare the following methods for inference: (i) the kernel estimator of the pointwise variance (\ref{eq:point_se}), which can be computed very quickly, (ii) the  empirical bootstrap of the quantile regression coefficients that solves the full optimization problem (\ref{eq:qr_estimator}) and is the most demanding in term of computation time, (iii) the empirical bootstrap   of the one-step estimator (\ref{alg:onestep}), which uses a linear approximation in the quantile but not in the bootstrap direction, (iv) the score multiplier bootstrap based on the quantile regression estimator, which uses a linear approximation in the bootstrap but not in the quantile direction, (v) the score multiplier bootstrap based on the one-step estimator, which uses two linear approximations and is the fastest bootstrap implementation. Note that we have initialized the one-step estimator at the median such that there is no difference at the median but there can be a difference at the $0.85$ quantile regression between the one-step and the original quantile regression estimators.

In small samples all methods over-reject the correct null hypothesis
but they all perform satisfactorily in large samples with empirical sizes that are very close to the theoretical size. The empirical bootstrap exhibits the lowest size distortion, which may warrant its higher computing burden. The analytic kernel-based method and the score multiplier bootstrap perform very similarly; this is not a surprise because they should provide the same standard errors when the number of bootstrap replications goes to infinity. Thus, there is no reason to use the score bootstrap when the goal is only to perform pointwise inference. The tests based on the one-step estimator displays a poor performance in very small samples. This is simply the consequence of the poor quality of the point estimates in very small samples that was shown in Table \ref{table:perf_one}. Like for the point estimates, the performance improves quickly and becomes almost as good as the tests based on the original quantile regression estimator.

\begin{table}
\begin{center}
\captionsetup{font=large}
\caption{Pointwise inference}
\label{table:pointwise}
\begin{tabular}{lcccccccc}
\hline \hline \noalign{\smallskip} 
 & \multicolumn{4}{c}{Median regression} & \multicolumn{4}{c}{0.85 quantile regression} \\
 & \multicolumn{4}{c}{\# of observations} & \multicolumn{4}{c}{\# of observations}\\
Inference method & 100 & 500 & 1000 & 5000 & 100 & 500 & 1000 & 5000\\
\noalign{\smallskip}\hline \hline \noalign{\smallskip}
\multicolumn{5}{c}{Empirical size (theoretical size 5\%) }\\
kernel & 0.12 & 0.08 & 0.07 & 0.05 & 0.13 & 0.08 & 0.07 & 0.05\\
empirical bootstrap & 0.07 & 0.06 & 0.06 & 0.05 & 0.09 & 0.07 & 0.07 & 0.05\\
empirical, one-step &  &  &  &  & 0.37 & 0.07 & 0.07 & 0.06\\
score bootstrap & 0.12 & 0.08 & 0.08 & 0.06 & 0.17 & 0.09 & 0.08 & 0.06\\
score, one-step &  &  &  &  & 0.34 & 0.10 & 0.08 & 0.06\\ & \multicolumn{4}{c}{Empirical power}\\
kernel & 0.50 & 0.98 & 1.00 & 1.00 & 0.34 & 0.92 & 1.00 & 1.00\\
empirical bootstrap & 0.49 & 0.99 & 1.00 & 1.00 & 0.35 & 0.94 & 1.00 & 1.00\\
empirical, one-step &  &  &  &  & 0.36 & 0.93 & 1.00 & 1.00\\
score bootstrap & 0.50 & 0.97 & 1.00 & 1.00 & 0.40 & 0.94 & 1.00 & 1.00\\
score, one-step &  &  &  &  & 0.34 & 0.91 & 1.00 & 1.00\\
\noalign{\smallskip}\hline\hline\end{tabular}\\
\end{center}
\end{table}
    
\subsection{Functional inference}
\label{sec:functional}

In this subsection we evaluate the performance of the implemented procedures for uniform inference. We consider the same data generating process (\ref{eq:dgp}) and test the  null hypotheses that the coefficient on $X$ is uniformly equal to $1$ (this is true) and that the coefficient on $X^2$ is uniformly equal to $0$ (this is false). We test these hypotheses with Kolmogorov-Smirnov (supremum of the deviations from the null hypothesis over the quantile range $\mathcal{T}$) and Cramer-von Mises statistics (average deviation from the null hypothesis over the quantile range $\mathcal{T}$), both with Anderson-Darling weights. We estimate the critical values using either the empirical bootstrap or the score bootstrap.\footnote{See \cite{ChFe05}, \cite{Angrist+06}, \cite{ChHa06} and \cite{belloni2017program} for more details and proofs of the validity of these procedures.}  We estimate the discretized quantile regression process for $\tau=0.1,0.11,0.12,...,0.9$ and we report the results for a theoretical size of 5\%.

Table \ref{table:uniform} provides the results. Unsurprisingly, the one-step estimator performs badly with $100$ observations. With this exception, all procedures perform satisfactorily for both test statistics even for moderate sample sizes. Even the fastest procedure (multiplier bootstrap based on the one-step estimators) is reliable when the sample size includes at least $500$ observations. This optimistic conclusion may be due to the low number of parameters ($k=3$) and to the smoothness of the conditional distribution of the response variable.. 

\begin{table}
\begin{center}
\captionsetup{font=large}
\caption{Uniform inference}
\label{table:uniform}
\begin{tabular}{lcccccccc}
\hline \hline \noalign{\smallskip} 
 & \multicolumn{4}{c}{Kolmogorov-Smirnov} & \multicolumn{4}{c}{Cramer-von-Mises} \\
 & \multicolumn{4}{c}{\# of observations} & \multicolumn{4}{c}{\# of observations}\\
Bootstrap method & 100 & 500 & 1000 & 5000 & 100 & 500 & 1000 & 5000\\
\noalign{\smallskip}\hline \hline \noalign{\smallskip}
 & \multicolumn{8}{c}{Empirical size}\\
empirical & 0.01 & 0.02 & 0.03 & 0.04 & 0.03 & 0.04 & 0.05 & 0.05\\
empirical, 1-step & 0.29 & 0.02 & 0.03 & 0.05 & 0.21 & 0.04 & 0.05 & 0.05\\
multiplier & 0.06 & 0.04 & 0.04 & 0.05 & 0.02 & 0.03 & 0.03 & 0.04\\
multiplier, 1-step & 0.45 & 0.06 & 0.05 & 0.06 & 0.09 & 0.03 & 0.03 & 0.04\\\noalign{\smallskip}\hline \noalign{\smallskip}
 & \multicolumn{8}{c}{Empirical power}\\
empirical & 0.92 & 1.00 & 1.00 & 1.00 & 0.94 & 1.00 & 1.00 & 1.00\\
empirical, 1-step & 0.58 & 1.00 & 1.00 & 1.00 & 0.60 & 1.00 & 1.00 & 1.00\\
multiplier & 1.00 & 1.00 & 1.00 & 1.00 & 0.98 & 1.00 & 1.00 & 1.00\\
multiplier, 1-step & 1.00 & 1.00 & 1.00 & 1.00 & 0.94 & 1.00 & 1.00 & 1.00\\
\noalign{\smallskip}\hline\hline\end{tabular}\\
\end{center}
\end{table}

\section{An empirical illustration}
\label{sec:application}

In this section we update the results obtained by \cite{abrevaya2001effects} and \cite{Koenker2001} using data from June 1997. They utilized quantile regression to analyze the effect of prenatal care visits, demographic characteristics and mother behavior on the distribution of birth weights. We use the 2017 Natality Data provided by the National Center for Health Statistics.  Like \cite{abrevaya2001effects} and \cite{Koenker2001} we keep only singleton births,  with mothers recorded as either black or white, between the
ages of 18 and 45, residing in the United States. We also dropped observations with missing data for
any of the regressors. This procedure results in a sample of 2,194,021 observations.

\begin{figure}
\begin{center}
\includegraphics[width=12cm]{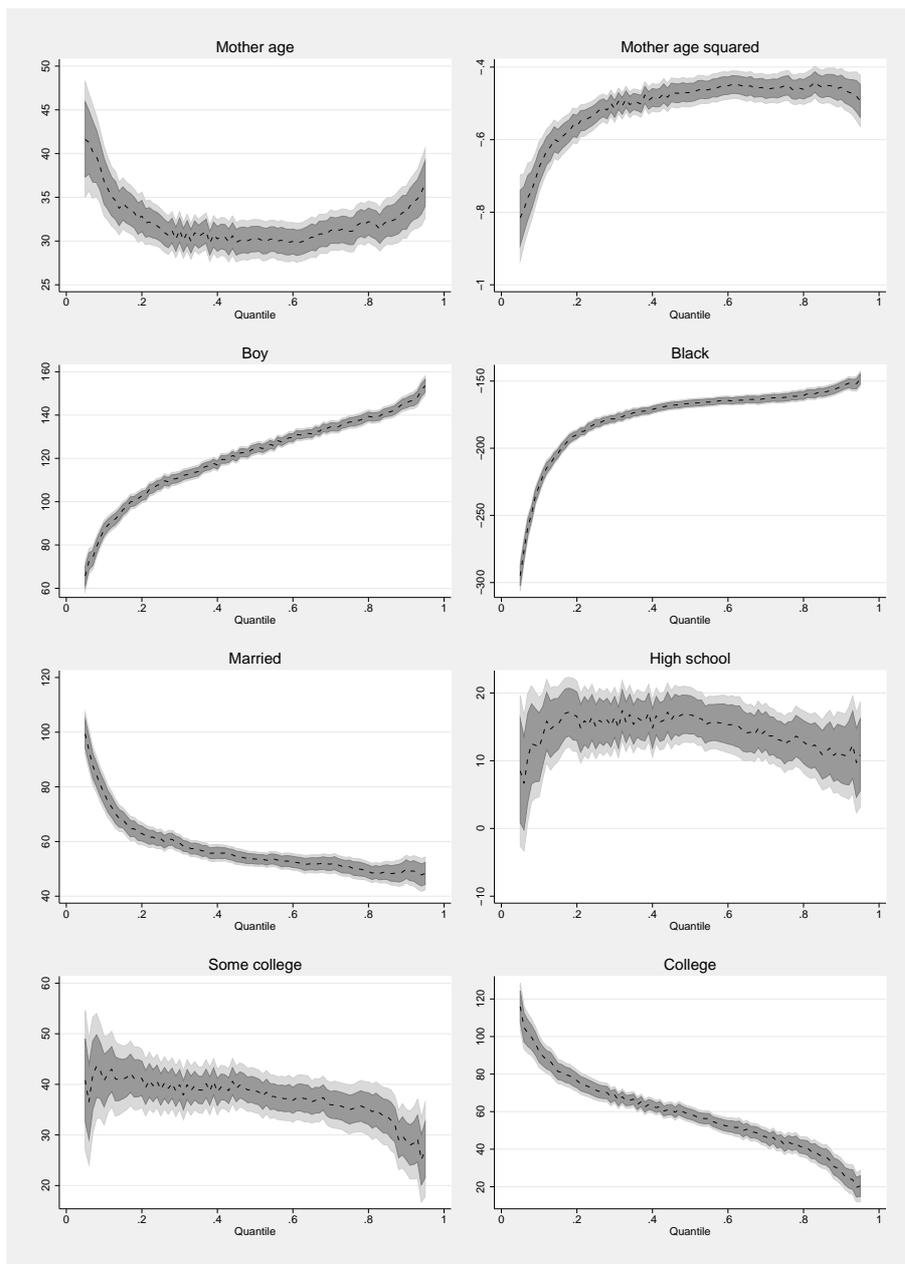}
\caption{Quantile regression estimates of the birth weight model}
\label{fig1}
\end{center}
\end{figure}

\begin{figure}
\begin{center}
\includegraphics[width=12cm]{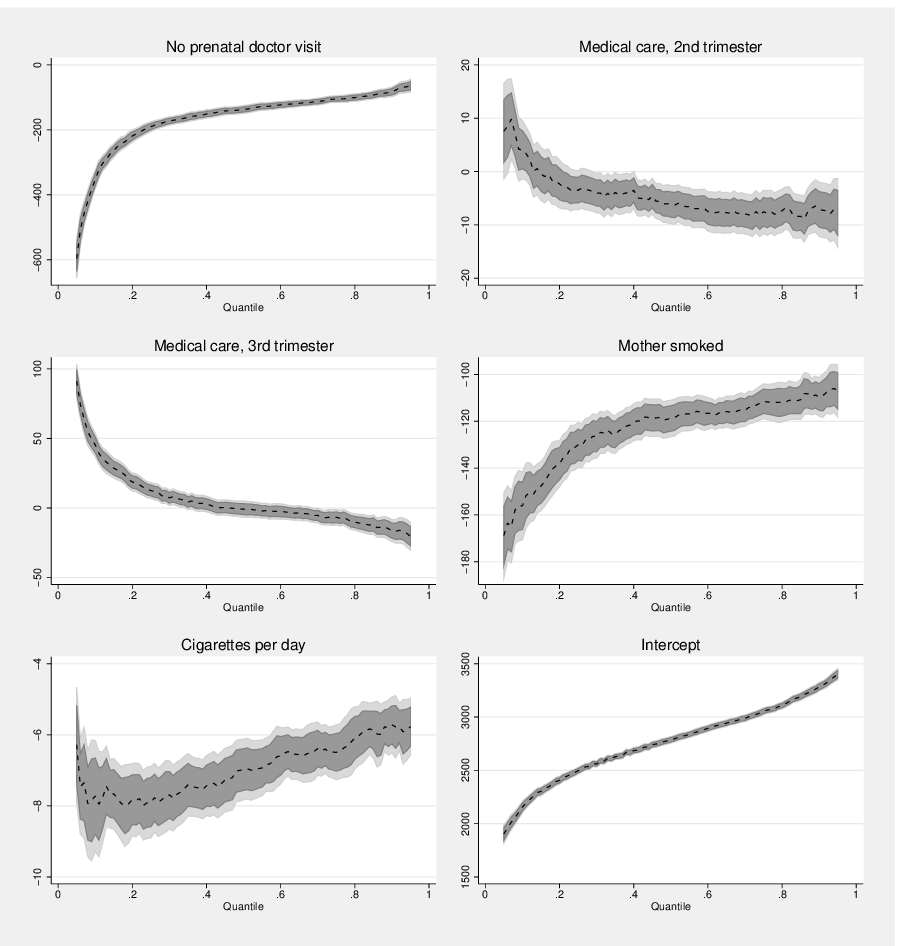}
\caption{Quantile regression estimates of the birth weight model (cont.)}
\label{fig2}
\end{center}
\end{figure}

We regress the newborn birth weights in grams on a constant and 13 covariates: gender of the newborn; race, marital status, education (4 categories), and age (and its square)  of the mother; an indicator for whether the mother smoked during the pregnancy and the average number of cigarettes smoked per day; and the prenatal medical care of the mother divided in four categories. We estimate 91 quantile regressions for $\tau=0.05,0.06,...,0.95$ and perform 100 bootstrap replications.\footnote{Due to computational limitations, \cite{abrevaya2001effects} and \cite{Koenker2001} used only the June subsample to estimate 5, respectively 15, different quantile regressions and they avoided bootstrapping  the results. Of course, computers have become more powerful in the meantime.} Given the large number of observations, regressors, quantile regressions and bootstrap replications, we use the fastest procedures, which is the one-step quantile regression estimator combined with the score multiplier bootstrap.
The computation of all these quantile regressions and bootstrap simulations took about 30 minutes on a 4-cores processor cloked at 3.7 GHz, i.e. a relatively standard notebook. With the built-in Stata commands the computation of the same quantile regressions and bootstrap should take more than 2 months. The new algorithms clearly open new opportunities for quantile regression.  

Figures \ref{fig1} and \ref{fig2} present the results for this example. For each coefficient, the line shows the point estimates, the dark grey area shows the 95\% pointwise confidence intervals and the light grey area shows the 95\% uniform confidence bands, which are wider by construction.\footnote{See the supplementary appendix SA to \cite{chernozhukov2013inference} for the construction and the validity of the uniform bands.} Any functional null hypothesis that lies outside of the uniform confidence band even at a single
quantile can be rejected at the 5\% level. For instance, it is obvious that none of the bands contains the value $0$ at all quantiles. We can, therefore, reject the null hypothesis that the regressor is irrelevant in explaining birth weight for all included covariates. For all variables we can also reject the null hypothesis that the quantile regression coefficient function is constant (location-shift model) because there is no value that is inside of the uniform bands at all quantiles.\footnote{The largest p-value is $0.02$ for high school.} On the other hand, we cannot reject the location-scale null hypothesis for some covariates because there exists a linear function of the quantile index that is covered by the band. This is the case for instance for all variables measuring the education of the mother.

Birth weight is often used as a measure of infant health outcomes at birth. Clearly, a low birth weight is associated 
 in the short-run with higher one-year mortality rates and in the longer run with worse educational attainment and
earnings, see, e.g., \cite{black2007cradle}. On the other hand, the goal of policy makers is clearly not to maximize birth weights. In a systematic review, \cite{baidal2016risk} find that higher birth weight was
consistently associated with higher risk of childhood
obesity. Thus, we do not want to report the average effects   of potential determinants of the birth weight but we need to analyze separately their effects on both tails of the conditional birth weight distribution. In the location-shift model, the effect of the covariates is restricted to be the same over the whole distribution and it can be estimated by least-squares methods. But the homogeneity of the effects is rejected for all regressors such that we need quantile regression.   

 In light of this discussion, it is interesting to see that the effect of a higher mother education (especially of a college degree) is much higher at the lower end of the distribution where we want to increase the birth weight. Similarly, not seeing a doctor during the whole pregnancy has dramatic effect at the lower end of the distribution. While not recommended,  not seeing a doctor has only mild effect at the top of the distribution.   In general, the results are surprisingly similar to the results reported by \cite{Koenker2001} in their Figure 4. Twenty years later, the shape, the sign and even the magnitude of most coefficients are still similar.

\section{Prospects}
\label{sec:prospects}

The advance of technology is leading to the collection of such a large amount of data that they cannot be saved or processed on a single computer. When this is the case,  
\cite{volgushev2019distributed} suggest estimating separately $J$ quantile regressions on each computer and averaging these estimates. In a second step, they obtain the whole quantile process by using a B-spline interpolation of the estimated quantile regressions.  We think that the new algorithms, in particular the one-step estimator, can be useful for the distributed computation of the quantile regression process. The computation of the one-step estimator requires only the communication of a $k\times1$ vector and a $k\times k$ matrix of sample means, which can be averaged by the master computer. Thus, the one-step estimator can be implemented even when the data set must be distributed on several computers.

The optimization problem that defines the linear quantile regression estimator is actually easy to solve because it is convex. The optimization problems defining, for instance, censored quantile regression, instrumental variable quantile regression, binary quantile regression are more difficult to solve by several orders of magnitude because they are not convex.\footnote{See \cite{Po87} for the censored quantile regression estimator, \cite{ChHa06} for the instrumental variable quantile regression estimator and \cite{kordas2006smoothed} for the binary quantile regression estimator, which is a generalization of \cite{Ma75} maximum score estimator.} Estimating the whole quantile process for these models may not be computationally feasible for the moment. It should be possible to adapt the suggested preprocessing and one-step algorithms to estimate the quantile processes also in these models.


%
%

\begin{acknowledgements}
We would like to thank the associate editor Roger Koenker, two annomynous referees, and the participants to the conference ``Economic Applications of Quantile Regressions 2.0'' that took place at the Nova School of Business and Economics for useful comments.
\end{acknowledgements}

%
%

\bibliographystyle{spbasic}      
\bibliography{research}   

\end{document}